\documentclass[%
aps,
amsmath,amssymb,
reprint,%
]{revtex4-1}

\usepackage{graphicx}
\usepackage{dcolumn}
\usepackage{bm}

\usepackage[utf8]{inputenc}
\usepackage[T1]{fontenc}
\usepackage{mathptmx}
\usepackage{color}

\begin{document}

	
	\title{On the micromagnetic behavior of dipolar-coupled nanomagnets in defective square artificial spin ice systems}
	
	\author{Neeti Keswani}
	\author{Pintu Das}%
	\email{pintu@physics.iitd.ac.in}
	\affiliation{ Department of Physics, Indian Institute of Technology, Delhi, Hauz Khas, New Delhi-110016, India}

	\date{\today}
	
	\begin{abstract}
		
		We report here the results of micromagnetic simulations of square artificial spin ice (ASI) systems with defects. The defects are introduced by misaligning of a nanomagnet at the vertex. In these defective systems, we are able to stabilize emergent monopole-like state by applying a small external field. We observe a systematic change of dipolar energies of the systems with varying misalignment angle. The fields at which the emergent monopoles are created vary linearly with the dipolar energies of the systems. Our results clearly show that the magnetization reversal of the ASI systems is intricately related to the interplay of defects and dipolar interactions. 
	\end{abstract}

	\maketitle

\begin{figure}
		\includegraphics[width=1\linewidth]{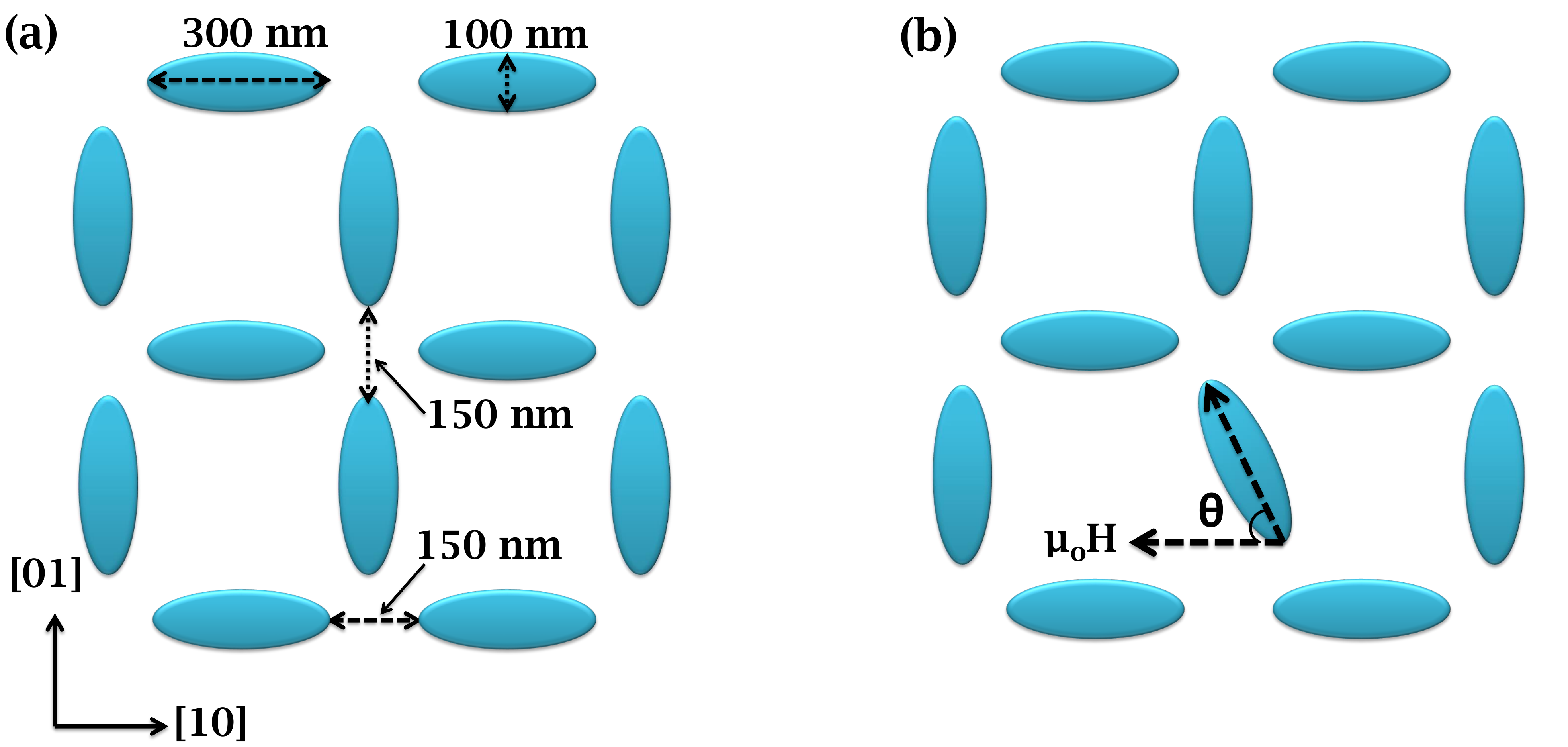}
		\caption{\label{schematics} Schematics of a regular (without defect) vertex in square ASI geometry (a) and a defective vertex with misaligned nanoisland (b).}
	\end{figure}

	Artificial spin-ice (ASI) systems are lithographically patterned arrangements of interacting magnetic nanostructures that were introduced for investigating the effects of geometric frustration in a controlled manner\cite{wang2006artificial}. In particular, nanomagnets in an artificial 2D square and kagome array that mimic the spin ice behavior have emerged as subject for intensive investigation in recent years\cite{wang2006artificial, ladak2011direct, gilbert2014emergent, libal2018ice, farhan2019emergent}. Square ASI can be considered as composed of two orthogonal sublattices of identical nanomagnets owing to their easy axes aligned along [10] and [01] directions. The recent progress in nano-lithography techniques enables us to  tune various parameters such as interaction strength between nanomagnets, their geometry as well as introduction of artificial defects\cite{silva2013nambu, budrikis2011diversity, montoncello2018mutual, gonccalves2019tuning}. In 2008,  Castelnovo \textit{et al.} realized that excitations above the degenerate ground states in spin ice systems, where ice rule is violated, could be interpreted as emergent magnetically charged quasiparticles that behave like magnetic monopoles\cite{castelnovo2008magnetic}. An important aspect of vertex frustration is its fascinating relation to the lattice topology and defects\cite{gilbert2016emergent, drisko2017topological, morrison2013unhappy, gilbert2014emergent, lao2018classical, farhan2013exploring}. This makes it possible to tune the complex dynamics of the magnetically charged vertices \cite{ladak2011direct,ladak2012disorder,gilbert2016emergent, wang2016rewritable, gartside2018realization, chavez2018voltage, gonccalves2019tuning}.  During fabrication of a large array of such nanostructures, it may be possible that a nanoisland which can be considered as a macrospin may be misaligned or it may loose its magnetically single-domain character due to an unintentional structural defect occuring during the fabrication steps. In artificial spin ice systems, the impact of such defects in the overall spin ice behavior can be studied in a controlled way. Moreover, the creation or annihilation of excited states are connected to the magnetization reversal of the nanostructures at the vertex. In order to understand this aspect of ASI systems, we carried out detailed micromagnetic simulation studies for individual square-ASI vertices. In a recent paper, we reported the observation of stable emergent monopole-like state in an even numbered vertex with vacancies at specific square lattice sites at the edges~\cite{keswani2018magnetization}. In this paper, we report the energetics of individual vertices as defects are introduced in the form of controlled misalignment of a vertex island. The magnetization reversals of the magnetic nanostructures in the form of elliptical shaped nanoislands of the vertices exhibit an angle dependent behavior. An emergent monopole-like state was stabilized in these structure with defects. 
		The schematics of the ASI structure used for the simulations is shown in Fig.\,\ref{schematics}(b). 
	
	Elliptical nanomagnets of Ni$_{80}$Fe$_{20}$ of aspect ratio 3 with dimensions of $300\times100\times25$\,nm$^3$ were used in square ASI geometry. The lattice constant for the lattice consisting of these elliptical nanomagnets is defined as the separation between edge-to-edge of the nanomagnets. The lattice constant is 150\,nm (see Fig.\,\ref{schematics}(a)).  For micromagnetic simulations of the individual vertices, finite difference based three dimensional solver of the Object Oriented Micro Magnetic Framework (OOMMF), which is an openware software from the National Institute of Standards and Technology, is used \cite{donahue1999national}. The time dependent magnetization dynamics in the nanoisland is governed by the Landau-Lifshitz-Gilbert (LLG) equation,
	\begin{center}
		
		$\frac{d\textbf{m}}{dt}$ = -$\mid$$\gamma$$\mid$\textbf{m} $\times$ \textbf{H}$_{eff}$ + $\alpha$(\textbf{m} $\times$ $\frac{d\textbf{m}}{dt})$
	\end{center}
	where $\gamma$ is the gyromagnetic ratio, and $\alpha$ is the Gilbert damping constant. The effective magnetic field is given by \textbf{H}$_{eff}$ = -$\frac{1}{\mu_o}$  ($\delta$W/$\delta$m), 
	where $\mu_o$ is the vacuum permeability, and W is the magnetic energy of the system which  consists of exchange, anisotropy and dipolar energy. For simulations, the nanoislands are discretized into cubic cells with each side of dimension 5\,nm, which is less than the exchange length ($\sim$\,5.3\,nm). For the calculations, typical experimentally reported values of saturation magnetization $M_s = 8.6 \times10^5$\,A/m, the exchange stiffness constant $A = $13\,pJ/m and damping constant of 0.5 for Ni$_{80}$Fe$_{20}$ are used\cite{principles}. The anisotropy is dominated by the shape ($K_{\rm{shape}}\approx 7.3\times10^4$Jm$^{-3}$) of the nanomagnets and therefore, the magnetocrystalline anisotropy is neglected in the computation. For these dimensions, the nanomagnets are magnetically in single domain state which was verified from simulations as well as experiments (not shown). Starting from a randomized state, the system was allowed to reach its minimum energy. Thereafter, the state was saturated by applying a magnetic field of 200\,mT along [10] direction. The magnetization reversal was studied while sweeping the field between $\pm\,200$\,mT at \textit{T} = 0\,K. 
	
\begin{figure}[h]
	\includegraphics[width=1\linewidth]{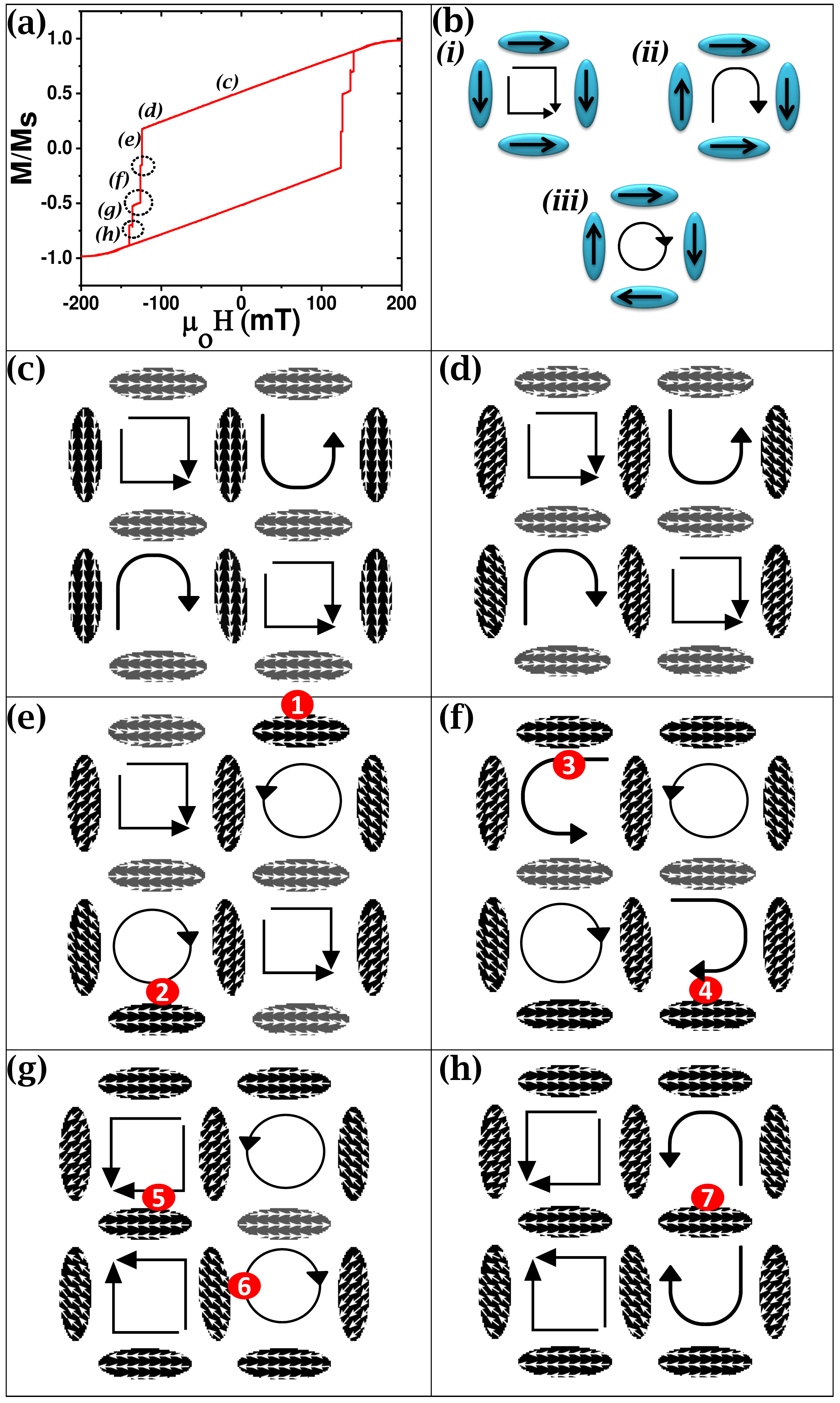}
	\caption{\label{5deg}(a) Hysteresis loop for a vertex with closed edges for misaligned angle $\theta$ = $85^\circ$. (b) Schematics of (i) onion state (ii) horse-shoe state and (iii) micro-vortex state.  (c)-(h) Magnetic states at intermediate field showing magnetic switchings as shown in (a).}
\end{figure}

	In the earlier work of magnetization reversal behavior of a regular vertex with closed edges, it was observed that the remanent state follows a two-in/two-out spin ice state\cite{keswani2018magnetization}. Detailed analysis of the micromagnetic behavior showed the specific way in which the reversal of the vertex magnetization proceeds. Here a deformity is introduced by misaligning the easy axis of one of the vertex-nanoislands in a sublattice with respect to the applied magnetic field direction. We define the angle between the misaligned easy axis and the applied field direction as misalignment angle $\theta$ (see Fig.\,\ref{schematics}(b)). If such a misalignment (defect) exists in a large array of square ASI system, the defect may have a significant role in overall behavior of the array. Therefore, primarily in this work we simulate such individual defective vertex structures and carry out systematic investigations of their micromagnetic behavior for  varying misalignment angle $\theta$ where, 20$^{\circ} \leq \theta \leq 85^{\circ}$. For the nanomagnet under consideration, $\theta = 90^\circ$ corresponds to the applied field direction oriented along its hard axis. Thus, for smaller values of $\theta$, the orientation of the field approaches towards easy direction of the nanomagnet. 
	Initially at $\theta$ = $85^\circ$, the system is saturated by applying a positive magnetic field of 200\,mT along [10] direction as indicated in Fig.\,\ref{5deg}(a). Thereafter while sweeping the field between $\pm200$\,mT, the micromagnetic calculations are carried out at different external fields.  
	The static equilibrium magnetization at remanence evolves to a two-in/two-out magnetic state as depicted in Fig.\,\ref{5deg}(c). The edge magnetic configurations at remanence and beyond are identified as consisting of two onion and two horse-shoe type chiral states as illustrated in Figs.\,\ref{5deg}(c) and (d), respectively. The onion state refers to two sublattices of nanoislands with parallelly aligned magnetization whereas horse-shoe state corresponds to parallelly aligned magnetization of one sublattice and antiparallelly aligned magnetization of the other sublattice as illustrated in Fig.\,\ref{5deg}b(i) and (ii). On the otherhand, the micro-vortex state consists of sublattices of nanoislands with opposite magnetizations\cite{remhof2008magnetostatic,keswani2018magnetization} (see Fig.\,\ref{5deg}b(iii)). Considering the macro-spin model\cite{domains} for these single domain nanomagnetic islands, four head-to-tail magnetic configurations form a micro-vortex loop whereas both onion and horse-shoe states have two head-to-tail, one head-to-head and one tail-to-tail configurations (see Fig.\,\ref{5deg}(b)). Calculations based on macro-spin model show that the energy hierarchy of these states follows $E_{microvortex}<E_{horse-shoe}<E_{onion}$\cite{keswani2018magnetization}. As shown in Fig.\,\ref{5deg}(a), the magnetization of the system is found to reverse in four distinct steps where sharp jumps are observed within a small field range of 124\,mT to 140\,mT. To understand the origin of these sharp jumps, we investigated the exact micromagnetic states of the system at every 2\,mT during the reversal. Fig.\,\ref{5deg}(d) shows the magnetic configuration just before the first jump (hereafter switching), showing two-in/two-out magnetic configuration at the vertex. It is observed that the first sharp jump at $\mu_o H = -124$\,mT corresponds to the simultaneous reversals (switching) of magnetization of two nanomagnets 1 and 2 situated at the diagonally opposite positions at 1$^{st}$ and 3$^{rd}$ quadrants (see Fig.\,\ref{5deg}(e)). These reversals convert the two horse-shoe type loops to two lower energy micro-vortex loops.
	The next jump in hysteresis occurs at $\mu_o H$ = -126\,mT, where the other diagonally opposite nanomagnets 3 and 4 switch simultaneously thereby converting two onion states to two lower-energy horse-shoe states as shown in Fig.\,\ref{5deg}(f). As the field is further increased in the negative direction, the nanomagnets 5 and 6 in two different sublattices switch simultaneously (see Fig.\,\ref{5deg}(g)). Interestingly, we observe that due to the misalignment angle of 85$^{\circ}$, a small component of the field ($~\mu_0H\cos\theta$) along the easy axis direction of the nanomagnet assists the switching of magnetization of the nanomagnet to one of its minimum energy states.
Such switching in the similarly positioned nanomagnets (i.e., easy axis orthogonal to the applied field) in the non defective ASI vertices were not observed before\cite{keswani2018magnetization}. This clearly demonstrates the tunability of the switching behavior by introducing such defects in ASI vertices. This also leads to the creation of a type-I state at the vertex. With further increase in field, the nanomagnet 7 finally reverses at $\mu_o H$ = -140\,mT. As shown in Fig.\,\ref{5deg}(h), this final reversal changes the vertex from two-in/two-out state to three-in/one-out state. According to the dumbbell model proposed by Castelnovo \textit{et al.}\cite{castelnovo2008magnetic}, a magnetic dipole (macro-spin) can be assumed to represent magnetic charges of +$Q_{\rm{m}}$ and -$Q_{\rm{m}}$. Thus, the vertex as shown in Fig.\,\ref{5deg}(h) has charge +2$Q_{\rm{m}}$ magnetic state which is an emergent magnetic monopole state. Remarkably, the emergent monopole-like state during reversal has not been observed for a regular vertex (without defects) with closed edges\cite{keswani2018magnetization} so far. Though the misaligned defect does not modify the remanent state of the vertex, it leads to a drastic change in reversal mechanism with the creation of emergent monopole-like (type-III state) from the lowest energy type-I state. Interestingly, the edge loops now changes to two onion and two horse-shoe type loops.

\begin{figure}[h]
	\includegraphics[width=1\linewidth]{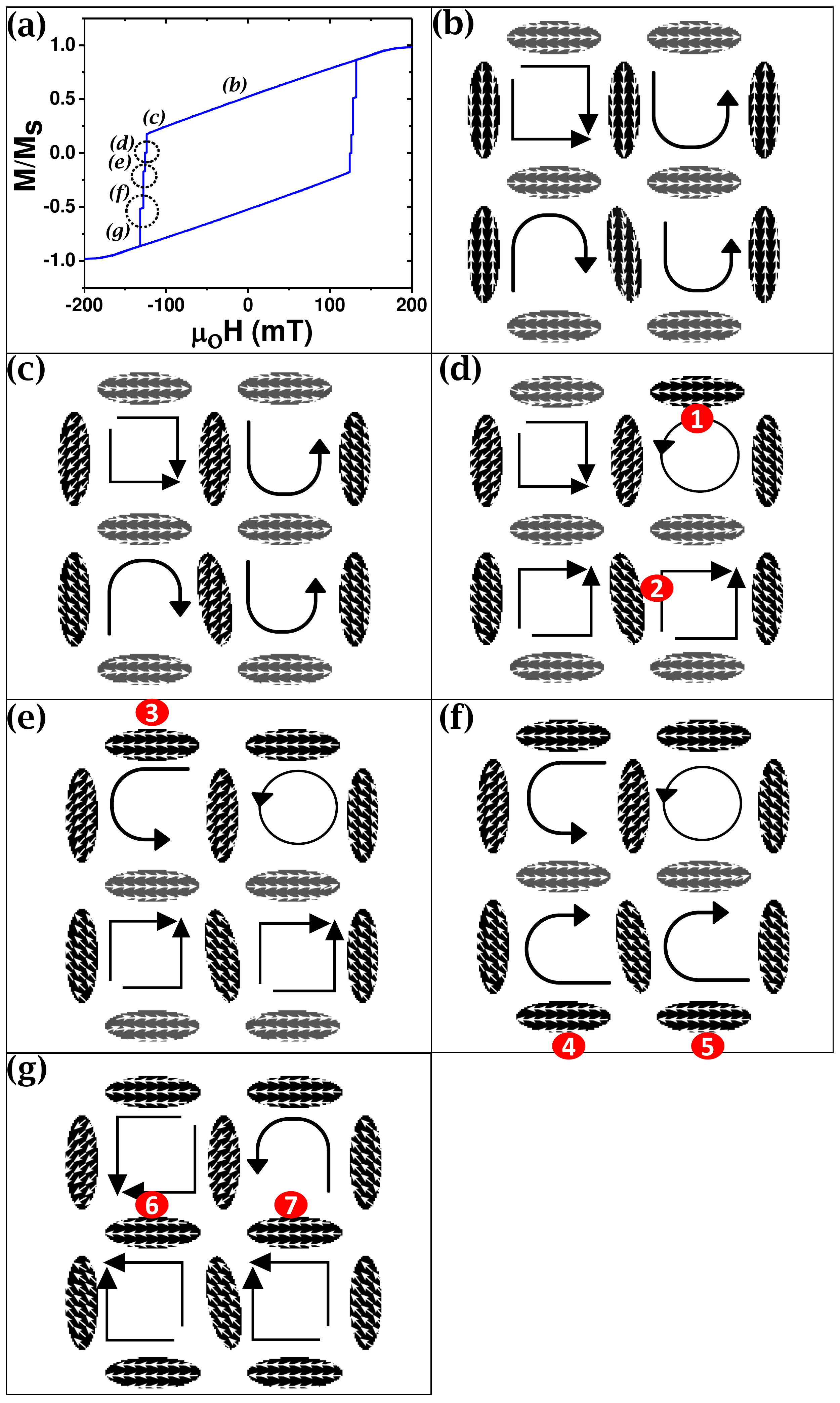}	
	\caption{\label{10deg}(a) Hysteresis loops for a system with misaligned angle $\theta$ = $80^\circ$. (b)-(g) Magnetic states at intermediate field showing magnetic switchings as shown in (a).}
\end{figure}

As the misalignment angle is changed from 85$^\circ$ to 80$^\circ$, we observe again a two-in/two-out (type-II) spin ice state at remanence as shown in Fig.\,\ref{10deg}(b). After saturation, the first sharp jump in the hysteresis is observed at $\mu_o H$ = -124\,mT as shown in Fig.\,\ref{10deg}(a). Simulation results show that this corresponds to simultaneous reversal of nanomagnets 1 and 2 of different sublattices (see Fig.\,\ref{10deg}(d)). In this case, the field has a stronger component along the easy axis direction of the misaligned nanomagnet. Thus, this nanomagnet switches at a relatively smaller field. This reversal converts one horse-shoe loop to one microvortex and the other horse-shoe to an onion type edge-loops. Furthermore, the vertex now turns to an emergent monopole-like state with magnetic charge +2$Q_{\rm{m}}$. With further increase in Zeeman energy, the magetization of nanomagnet 3 reverses so that a higher energy onion state is converted to a lower energy horse-shoe state at $\mu_oH$ = -126\,mT (see Fig.\,\ref{10deg}(e)). The next jump is observed at $\mu_o H$ = -128\,mT which corresponds to the simultaneous switchings of nanomagnets 4 and 5 thereby converting the two onion loops to two horse-shoe type loops. The final switching takes place at $\mu_o H$ = -132\,mT where the two nanomagnets at the vertex, indicated by 6 and 7, reverse simultaneously as shown in Fig.\,\ref{10deg}(g). It is remarkable that the vertex remains at the +2$Q_{\rm{m}}$ charged state during the entire reversal process. Thus a stable emergent magnetic monopole state is generated for this defect configuration.

\begin{figure}
	\includegraphics[width=1\linewidth]{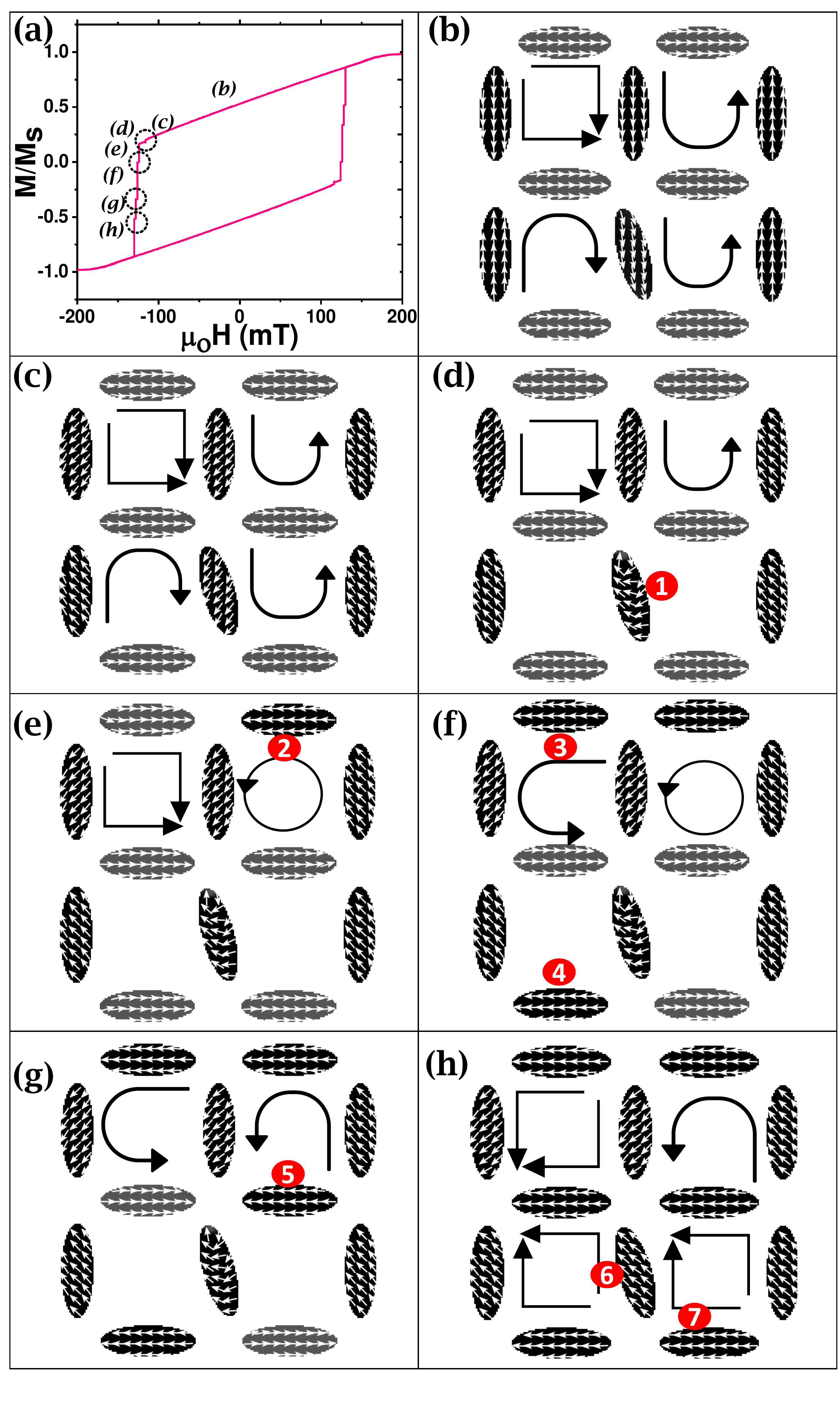}
	\caption{\label{15deg}(a) Hysteresis loops for a vertex with closed edge with misalignment angle of $75^\circ$. (b)-(h) Magnetic states after corresponding magnetic switching of individual nanomagnets  as shown in (a).}
\end{figure}

With further reduction of the misalignment angle by 5$^\circ$ i.e., for $\theta = 75^\circ$, interesting changes are observed in the micromagnetic behavior of the nanomagnets. The hysteresis shows five distinct jumps for $\theta = 75^\circ$ as seen in Fig.\,\ref{15deg}(a). The remanent state still follows type-II spin ice state as for the other cases (see Fig.\,\ref{15deg}(b)). 
The first jump in the hysteresis loop - which takes place at lower field $\mu_o H$ = -116\,mT - corresponds to a change in the detailed micromagnetic pattern in the misaligned nanomagnet 1 from single domain to magnetic vortex state. The core of the magnetic vortex lies at the edge of the elliptical nanomagnet which is close to the ASI vertex as shown in Fig.\,\ref{15deg}(d). The chirality of magnetic vortex in this nanomagnet appears to depend on the direction of the field sweep. Vortex of clockwise chirality was observed while downsweeping the field from 200\,mT to -200\,mT as shown in Fig.\,\ref{15deg}(d-f). Vortex of opposite chirality is observed while upsweeping the field (not shown). Such magnetic vortex structure is purely due to the energy minimization as a result of a complex interplay of the defect, dipolar interactions and the magnetic field. The change of single domain state to a magnetic vortex state is a significant result which clearly underlines the role of the defect in creating new states in the nanomagnets. As a result of this magnetic vortex-type magnetic structure, the two-in/two-out spin ice rule at the vertex is violated, thereby converting the vertex to an anomalous state where the nanomagnet does not retain the single domain behavior. Considering the magnetic vortex in the nanomagnet as magnetically chargeless, the net charge at this vertex is +$Q_{\rm{m}}$. 
As the negative field is further increased, the second jump is observed at $\mu_o H$ = -124\,mT which corresponds to switching of nanomagnet 2 (see Fig.\,\ref{15deg}(e)). The next jump is observed at $\mu_o H$ = -126\,mT, where the magnetization of the two nanomagnets 3 and 4 reverses simultaneously. This is shown in Fig.\,\ref{15deg}(f). The fourth and fifth jumps occur at $\mu_o H$ = -128\,mT and $\mu_o H$ = -130\,mT, respectively. The jump at $\mu_o H$ = -128\,mT is due to the reversal of magnetization of nanomagnet 5 as shown in Fig.\,\ref{15deg}(g). With the reversal of nanomagnet 5, the charge of 3$Q_{\rm{m}}$ is generated at vertex. This additional magnetic charge observed at the vertex is due to the creation of a magnetic vortex in the misaligned nanomagnet.
The jump at $\mu_o H$ = -130\,mT corresponds to the simultaneous reversals of magnetizations of nanomagnets 6 and 7, respectively as illustrated in Fig.\,\ref{15deg}(h).

\begin{figure}[h]
	\includegraphics[width=1\linewidth]{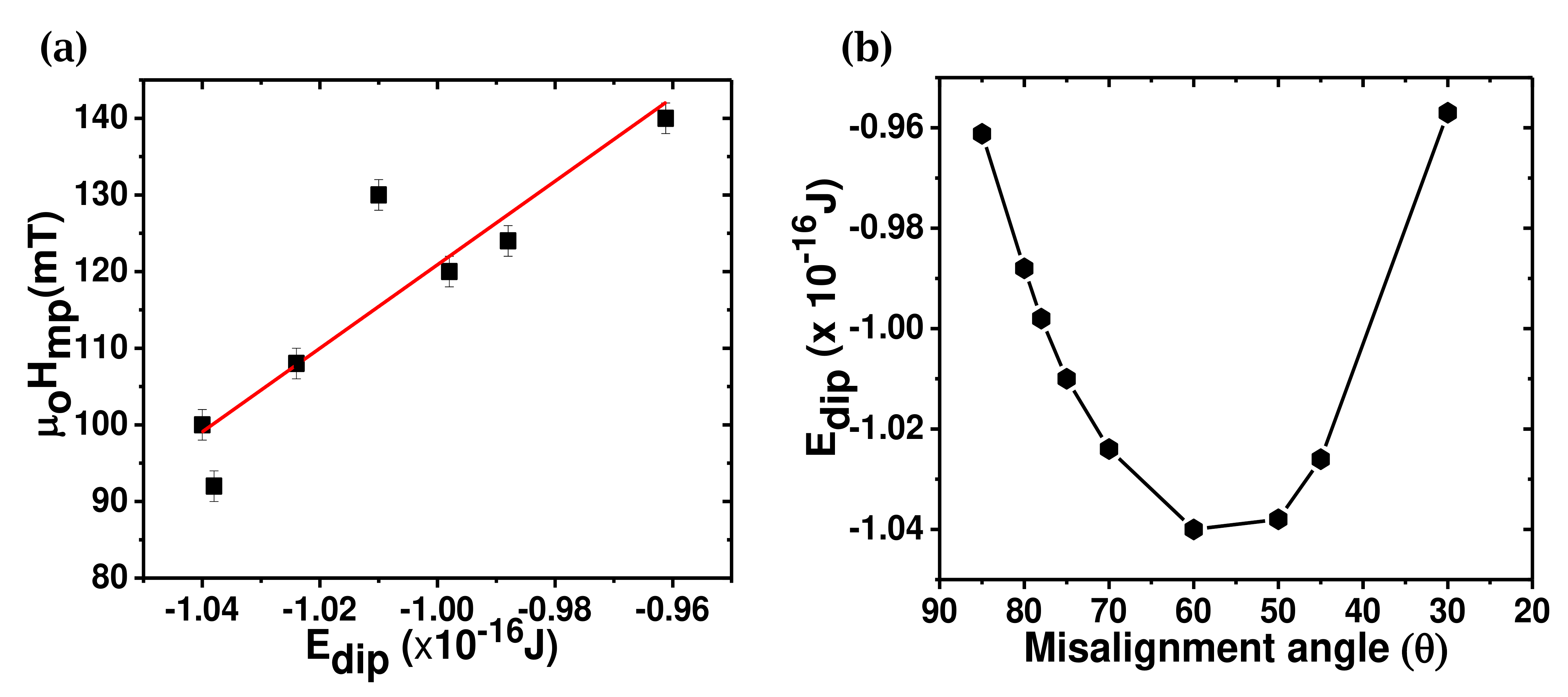}
	\caption{\label{graph}(a) Field ($\mu_0H_{\rm{mp}}$) at which monopole is generated vs. the dipolar energy (E$_{dip}$) of the system. (b) Dipolar energy ($E_{\rm{dip}}$)  vs. the angle of misalignment ($\theta$) in the misaligned vertex.}
\end{figure}

Interestingly, with the final reversals, the magnetic state of the misaligned nanomagnet again changes back to a single domain state thereby creating a charge of +2$Q_{\rm{m}}$, i.e., an emergent monopole-like state at the vertex. After complete reversal of nanomagnets, the magnetic state constitutes one horse-shoe and three onion states (see Fig.\ref{15deg}(h)). Thus, for $\theta = 75^{\circ}$ we observe the creation of an emergent monopole-like state with three-in/one-out magnetic orientation at the vertex via an anomalous state (of charges $Q_{\rm{m}}$ and 3$Q_{\rm{m}}$). When carefully calculated, we find that such vortex type structure in the misaligned nanomagnet appears for angle atleast till $\theta = 73^{\circ}$ however, it disappears for $\theta \leq70^{\circ}$. The calculations are performed for various angles till $\theta = 20^\circ$. In general, for $\theta \leq70^{\circ}$, the reversal mechanism exhibits similar behaviour as described for misalignment angle $\theta = 80^{\circ}$. The emergence of monopole-like state appears for all the misalignment angles studied in this work. It is clear that such emergent monopole-like magnetically charged states can be predictably created at the vertices due to the interplay of defects and dipolar interactions. The corresponding magnetic states for all the misalignment angles studied in this work are summarised in Table 1.
To investigate the role of such defects in stabilizing a magnetic state, we plot the fields at which emergent monopoles are observed ($\mu_0H_{\rm{mp}}$) for each defective structures against the dipolar energy ($E_{\rm{dip}}$) of the respective structures (i.e., for different misalignment angles).  The linear dependence as shown in Fig.\,\ref{graph}(a) underlines the role of the interplay of defects and dipolar interactions in stabilizing the charged states. Fig.\,\ref{graph}(b) shows the dependence of dipolar energy ($E_{\rm{dip}}$) of the system on misalignment angle $\theta$. The dipolar energies are calculated using macro-spin model.

\begin{table}
	\centering
	
\begin{tabular}{ |p{2.1cm}|p{2.5cm}|p{4cm}|  }
	
	\hline
	\textbf{Misalignment Angle} & \textbf{Remanent State} &  \textbf{Magnetic states at intermediate field}\\
	\hline
	$85^{\circ}$  &   Type II state & Type I (136\,mT)  and Type III (140\,mT) \\
	\hline
	$80^{\circ}$ &  Type II state &Type III (124\,mT) \\
	\hline
	$75^{\circ}$ &  Type II state &undefined state (116\,mT) and Type III (128\,mT)\\
	\hline
	$50^{\circ}$ &   Type II state  &Type III(92\,mT)\\
	\hline
	$45^{\circ}$ &   Type II state  &Type III(88\,mT)\\
	\hline
	$30^{\circ}$ &   Type II state  &Type III(90\,mT)\\
	\hline
	$20^{\circ}$ &   Type II state  &Type III(98\,mT)\\
	\hline
\end{tabular}
\caption{\label{Table}Table summaries the different misaligned angle and states generated at intermediate field. The numbers in bracket indicates the field at which the magnetic state is generated.}
\end{table}

In summary, we investigated in details the micromagnetic behavior of ASI vertices with defects in the form of a misaligned nanomagnet in the vertex. Systematic studies of the switching behavior exhibits sharp jumps in the hystersis loops. Detailed investigations of the jumps show that they correspond to switching of individual nanomagnets or simultaneous switchings of two nanomagnets in the ASI system. While switching, indirect couplings of two nanomagnets are observed in certain cases due to the strong dipolar interaction among the involved nanomagnets. For the misalignment angle of 75$^{\circ}$, an interesting change of the magnetic state of the misaligned nanomagnet from single domain to magnetic vortex state is observed. We observe a clear role of defects and dipolar interactions in stabilizing an emergent monopole-like state in such systems. Our results may allow design of artificial structures where defects and dipolar interactions can be used for specific device utilities. 
\begin{acknowledgments}
We acknowledge the High Performance Computing facilities (HPC) of IIT Delhi in performing the micromagnetic calculations. N.K is also thankful to University Grant Commision, (UGC) Govt. of India, for financial support. 
\end{acknowledgments}

\bibliography{NK_PD_arxiv}
	
\end{document}